\documentclass[runningheads]{llncs}
\usepackage{tikz}
\usetikzlibrary{3d,calc}
\usepackage{keyval}
\usepackage{ifthen}
\makeatletter
\newcommand{\tikzcuboid@shiftx}{0}
\newcommand{\tikzcuboid@shifty}{0}
\newcommand{\tikzcuboid@dimx}{3}
\newcommand{\tikzcuboid@dimy}{3}
\newcommand{\tikzcuboid@dimz}{3}
\newcommand{\tikzcuboid@scale}{1}
\newcommand{\tikzcuboid@densityx}{1}
\newcommand{\tikzcuboid@densityy}{1}
\newcommand{\tikzcuboid@densityz}{1}
\newcommand{\tikzcuboid@rotation}{0}
\newcommand{\tikzcuboid@anglex}{0}
\newcommand{\tikzcuboid@angley}{90}
\newcommand{\tikzcuboid@anglez}{225}
\newcommand{\tikzcuboid@scalex}{1}
\newcommand{\tikzcuboid@scaley}{1}
\newcommand{\tikzcuboid@scalez}{sqrt(0.5)}
\newcommand{\tikzcuboid@linefront}{black}
\newcommand{\tikzcuboid@linetop}{black}
\newcommand{\tikzcuboid@lineright}{black}
\newcommand{\tikzcuboid@fillfront}{white}
\newcommand{\tikzcuboid@filltop}{white}
\newcommand{\tikzcuboid@fillright}{white}
\newcommand{\tikzcuboid@shaded}{N}
\newcommand{\tikzcuboid@shadecolor}{black}
\newcommand{\tikzcuboid@shadeperc}{25}
\newcommand{\tikzcuboid@emphedge}{N}
\newcommand{\tikzcuboid@emphstyle}{thick}

\define@key{tikzcuboid}{shiftx}[\tikzcuboid@shiftx]{\renewcommand{\tikzcuboid@shiftx}{#1}}
\define@key{tikzcuboid}{shifty}[\tikzcuboid@shifty]{\renewcommand{\tikzcuboid@shifty}{#1}}
\define@key{tikzcuboid}{dimx}[\tikzcuboid@dimx]{\renewcommand{\tikzcuboid@dimx}{#1}}
\define@key{tikzcuboid}{dimy}[\tikzcuboid@dimy]{\renewcommand{\tikzcuboid@dimy}{#1}}
\define@key{tikzcuboid}{dimz}[\tikzcuboid@dimz]{\renewcommand{\tikzcuboid@dimz}{#1}}
\define@key{tikzcuboid}{scale}[\tikzcuboid@scale]{\renewcommand{\tikzcuboid@scale}{#1}}
\define@key{tikzcuboid}{densityx}[\tikzcuboid@densityx]{\renewcommand{\tikzcuboid@densityx}{#1}}
\define@key{tikzcuboid}{densityy}[\tikzcuboid@densityy]{\renewcommand{\tikzcuboid@densityy}{#1}}
\define@key{tikzcuboid}{densityz}[\tikzcuboid@densityz]{\renewcommand{\tikzcuboid@densityz}{#1}}
\define@key{tikzcuboid}{rotation}[\tikzcuboid@rotation]{\renewcommand{\tikzcuboid@rotation}{#1}}
\define@key{tikzcuboid}{anglex}[\tikzcuboid@anglex]{\renewcommand{\tikzcuboid@anglex}{#1}}
\define@key{tikzcuboid}{angley}[\tikzcuboid@angley]{\renewcommand{\tikzcuboid@angley}{#1}}
\define@key{tikzcuboid}{anglez}[\tikzcuboid@anglez]{\renewcommand{\tikzcuboid@anglez}{#1}}
\define@key{tikzcuboid}{scalex}[\tikzcuboid@scalex]{\renewcommand{\tikzcuboid@scalex}{#1}}
\define@key{tikzcuboid}{scaley}[\tikzcuboid@scaley]{\renewcommand{\tikzcuboid@scaley}{#1}}
\define@key{tikzcuboid}{scalez}[\tikzcuboid@scalez]{\renewcommand{\tikzcuboid@scalez}{#1}}
\define@key{tikzcuboid}{linefront}[\tikzcuboid@linefront]{\renewcommand{\tikzcuboid@linefront}{#1}}
\define@key{tikzcuboid}{linetop}[\tikzcuboid@linetop]{\renewcommand{\tikzcuboid@linetop}{#1}}
\define@key{tikzcuboid}{lineright}[\tikzcuboid@lineright]{\renewcommand{\tikzcuboid@lineright}{#1}}
\define@key{tikzcuboid}{fillfront}[\tikzcuboid@fillfront]{\renewcommand{\tikzcuboid@fillfront}{#1}}
\define@key{tikzcuboid}{filltop}[\tikzcuboid@filltop]{\renewcommand{\tikzcuboid@filltop}{#1}}
\define@key{tikzcuboid}{fillright}[\tikzcuboid@fillright]{\renewcommand{\tikzcuboid@fillright}{#1}}
\define@key{tikzcuboid}{shaded}[\tikzcuboid@shaded]{\renewcommand{\tikzcuboid@shaded}{#1}}
\define@key{tikzcuboid}{shadecolor}[\tikzcuboid@shadecolor]{\renewcommand{\tikzcuboid@shadecolor}{#1}}
\define@key{tikzcuboid}{shadeperc}[\tikzcuboid@shadeperc]{\renewcommand{\tikzcuboid@shadeperc}{#1}}
\define@key{tikzcuboid}{emphedge}[\tikzcuboid@emphedge]{\renewcommand{\tikzcuboid@emphedge}{#1}}
\define@key{tikzcuboid}{emphstyle}[\tikzcuboid@emphstyle]{\renewcommand{\tikzcuboid@emphstyle}{#1}}
\newcommand{\tikzcuboid}[1]{
    \setkeys{tikzcuboid}{#1} % Process Keys passed to command
    \pgfmathsetmacro{\vectorxx}{\tikzcuboid@scalex*cos(\tikzcuboid@anglex)}
    \pgfmathsetmacro{\vectorxy}{\tikzcuboid@scalex*sin(\tikzcuboid@anglex)}
    \pgfmathsetmacro{\vectoryx}{\tikzcuboid@scaley*cos(\tikzcuboid@angley)}
    \pgfmathsetmacro{\vectoryy}{\tikzcuboid@scaley*sin(\tikzcuboid@angley)}
    \pgfmathsetmacro{\vectorzx}{\tikzcuboid@scalez*cos(\tikzcuboid@anglez)}
    \pgfmathsetmacro{\vectorzy}{\tikzcuboid@scalez*sin(\tikzcuboid@anglez)}
    \begin{scope}[xshift=\tikzcuboid@shiftx, yshift=\tikzcuboid@shifty, scale=\tikzcuboid@scale, rotate=\tikzcuboid@rotation, x={(\vectorxx,\vectorxy)}, y={(\vectoryx,\vectoryy)}, z={(\vectorzx,\vectorzy)}]
    \pgfmathsetmacro{\steppingx}{1/\tikzcuboid@densityx}
    \pgfmathsetmacro{\steppingy}{1/\tikzcuboid@densityy}
    \pgfmathsetmacro{\steppingz}{1/\tikzcuboid@densityz}
    \newcommand{\dimx}{\tikzcuboid@dimx}
    \newcommand{\dimy}{\tikzcuboid@dimy}
    \newcommand{\dimz}{\tikzcuboid@dimz}
    \pgfmathsetmacro{\secondx}{2*\steppingx}
    \pgfmathsetmacro{\secondy}{2*\steppingy}
    \pgfmathsetmacro{\secondz}{2*\steppingz}
    \foreach \x in {\steppingx,\secondx,...,\dimx}
    {   \foreach \y in {\steppingy,\secondy,...,\dimy}
        {   \pgfmathsetmacro{\lowx}{(\x-\steppingx)}
            \pgfmathsetmacro{\lowy}{(\y-\steppingy)}
            \filldraw[fill=\tikzcuboid@fillfront,draw=\tikzcuboid@linefront] (\lowx,\lowy,\dimz) -- (\lowx,\y,\dimz) -- (\x,\y,\dimz) -- (\x,\lowy,\dimz) -- cycle;

        }
    }
    \foreach \x in {\steppingx,\secondx,...,\dimx}
    {   \foreach \z in {\steppingz,\secondz,...,\dimz}
        {   \pgfmathsetmacro{\lowx}{(\x-\steppingx)}
            \pgfmathsetmacro{\lowz}{(\z-\steppingz)}
            \filldraw[fill=\tikzcuboid@filltop,draw=\tikzcuboid@linetop] (\lowx,\dimy,\lowz) -- (\lowx,\dimy,\z) -- (\x,\dimy,\z) -- (\x,\dimy,\lowz) -- cycle;
        }
    }
    \foreach \y in {\steppingy,\secondy,...,\dimy}
    {   \foreach \z in {\steppingz,\secondz,...,\dimz}
        {   \pgfmathsetmacro{\lowy}{(\y-\steppingy)}
            \pgfmathsetmacro{\lowz}{(\z-\steppingz)}
            \filldraw[fill=\tikzcuboid@fillright,draw=\tikzcuboid@lineright] (\dimx,\lowy,\lowz) -- (\dimx,\lowy,\z) -- (\dimx,\y,\z) -- (\dimx,\y,\lowz) -- cycle;
        }
    }
    \ifthenelse{\equal{\tikzcuboid@emphedge}{Y}}%
        {\draw[\tikzcuboid@emphstyle](0,\dimy,0) -- (\dimx,\dimy,0) -- (\dimx,\dimy,\dimz) -- (0,\dimy,\dimz) -- cycle;%
        \draw[\tikzcuboid@emphstyle] (0,0,\dimz) -- (0,\dimy,\dimz) -- (\dimx,\dimy,\dimz) -- (\dimx,0,\dimz) -- cycle;%
        \draw[\tikzcuboid@emphstyle](\dimx,0,0) -- (\dimx,\dimy,0) -- (\dimx,\dimy,\dimz) -- (\dimx,0,\dimz) -- cycle;%
        }%
        {}
    \end{scope}
}

\usetikzlibrary{decorations.pathreplacing} % for curly braces

\usepackage{physics}

\colorlet{myblue}{blue!70!black}
\colorlet{mydarkblue}{blue!40!black}
\colorlet{mygreen}{green!40!black}
\colorlet{myred}{red!65!black}
\tikzstyle{vector}=[->,dashed,very thick,orange,line cap=round]
\tikzstyle{vector1}=[dashed,very thick,orange,line cap=round]
\tikzstyle{ptmiss}=[->,dashed,thick,myred,line cap=round]
\tikzstyle{cone}=[thin,blue!50!black,fill opacity=0.8]

\usetikzlibrary { decorations.pathmorphing, decorations.pathreplacing, decorations.shapes } 
\usepackage[T1]{fontenc}

\usepackage{graphicx}

\usepackage{float}
\usepackage{amsmath}
\usepackage{cleveref}

\begin{document}
\title{Towards Real Time Compton Imaging in Demanding Conditions}

\author{
Bernardo Gameiro\inst{1}\orcidID{0009-0000-7767-6410}\and
Jorge Lerendegui-Marco\inst{1}\orcidID{0000-0001-8358-9217}\and
Victor Babiano-Suarez\inst{2}\orcidID{0000-0002-2717-2123}\and\\
Javier Balibrea-Correa\inst{1}\orcidID{0000-0002-8404-3256}\and
Gabriel de la Fuente Rosales\inst{1}\orcidID{0000-0002-6452-9232}\and
Ion Ladarescu\inst{1}\orcidID{0000-0003-2405-281X}\and\\
Pablo Torres-Sánchez\inst{1}\orcidID{0000-0003-0608-5351}\and
César Domingo Pardo\inst{1}\orcidID{0000-0002-2915-5466}
}

\authorrunning{B. Gameiro et al.}

\institute{\inst{1}Instituto de Física Corpuscular, 46980 Paterna, Valencia, Spain\\
\email{\{bgameiro,Javier.Balibrea,Gabriel.delaFuente, 
Cesar.Domingo,Ion.Ladarescu,Jorge.Lerendegui,Pablo.Torres\}@ific.uv.es}\\
\inst{2}University of Valencia, 46100 Burjassot, Valencia, Spain\\
\email{Victor.Babiano@ific.uv.es}}
\maketitle              % typeset the header of the contribution
\vspace{-12pt}
\begin{abstract}
Compton cameras are radiation detectors that provide spatial information on the origin of the $\gamma$-ray sources based on the Compton scattering effect.

Many applications require these detectors to be used at high counting rate. As such, the preprocessing of the detections as well as the imaging algorithms are required to be time-efficient in order for the data to be processed in real time. 

In this work, optimizations to the preprocessing of events in Compton cameras based on monolithic crystals, with special focus on event identification, were implemented using a parallelizable algorithm.
Regarding imaging, an established 3D back projection algorithm was parallelized and implemented using SYCL \cite{SYCL}. The parallel implementation of the algorithm was included without and with several optimizations such as the pre-computing values, discarding low impact contributions based on angle, and selecting an efficient shape of the image universe. The implementations were tested with Intel CPUs, GPUs, and NVIDIA GPUs.

An outlook into the study of algorithms to reconstruct the position of interaction within Compton cameras based on monolithic crystals into segmented regions and other next steps is included.

\end{abstract}
\section{Introduction}
\subsection{Compton Imaging}

Compton imaging is a technique for imaging $\gamma$-rays based on Compton scattering, one of the mechanisms of interaction between high-energy photons and charged particles, such as electrons.

When a $\gamma$-ray interacts with a scintillator by Compton scattering, it deposits some of its energy, exciting electrons to a higher energy state. These electrons then deexcite releasing the excess energy by emitting photons. It is these photons that are collected by the photosensor associated with the scintillator to form the detector.

From the $\gamma$-ray energy before ($E_{\gamma }$) and after ($E_{\gamma^{\prime}}$) the interaction it is possible to calculate the scattering angle $\theta$:

$$E_{\gamma ^{\prime }}={\frac {E_{\gamma }}{1+(E_{\gamma }/m_{e}c^{2})(1-\cos \theta )}}$$

\begin{figure}[h!]
\centering
\includegraphics[width=.55\textwidth]{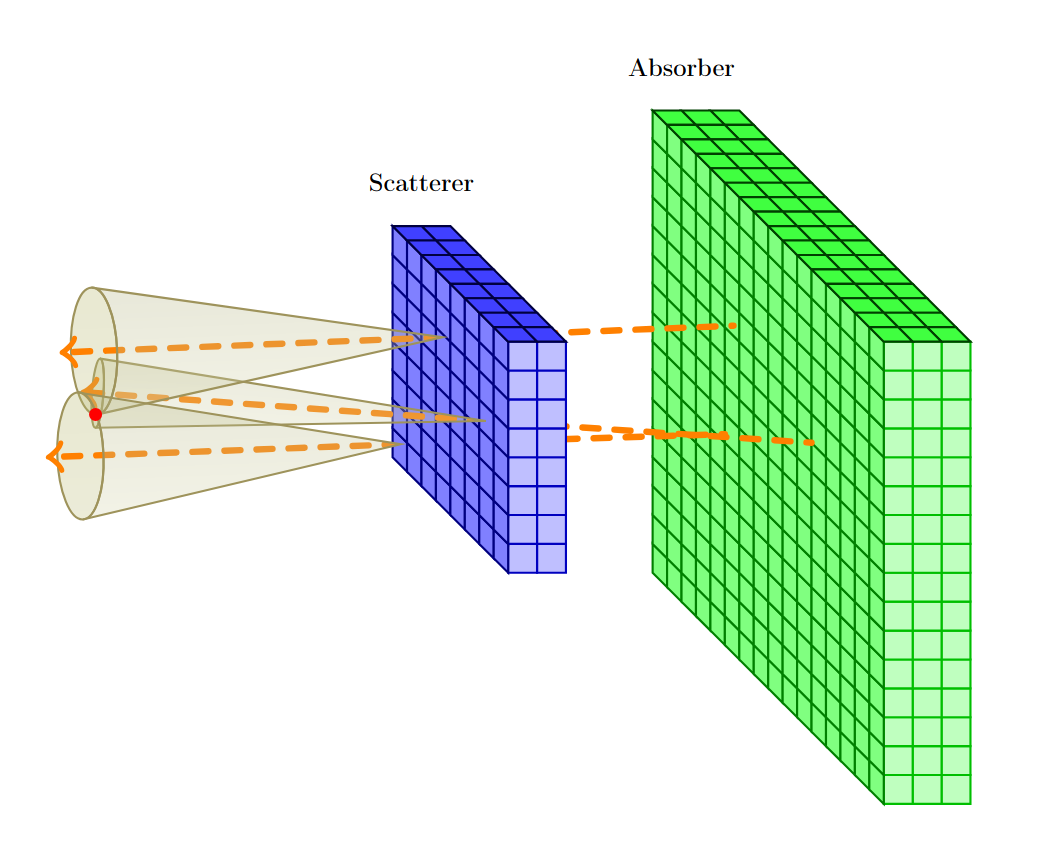}
\caption{Schematic showing a Compton camera setup with segmented crystals, cones defined from the measurements, and the origin of the $\gamma$-rays.} \label{ComptonCamera}
\end{figure}

By measuring the energy and position of the depositions in 2 detectors, one can draw the Compton cone, which consists on the possible positions of emission of the original $\gamma$-ray. By overlapping several cones, the position of origin can be computed, as seen in \cref{ComptonCamera}. 

%\section{Background}
\subsection{Compton Cameras}

Compton Cameras are a type of $\gamma$-ray detectors used in astrophysics \cite{HERZO1975583,nTOF_iTED}, medical imaging \cite{TODD,PT_iTED}, nuclear waste imaging \cite{GNVision} and homeland security. These detectors leverage the Compton Scattering effect in order to provide spatial information regarding the origin of $\gamma$ radiation. As such, these cameras inherently generate hyperspectral images, in contrast with PET-scanners, and involve physics computations, in contrast to Anger cameras, \cite{Anger}.

\subsubsection{Different Designs}

Although different designs of position-sensitive radiation detectors exist in order to address different requirements and applications, this work is going to focus on Compton cameras that use scintillators, the active material sensitive to radiation, coupled to a pixelated photosensor, which enable the position sensitivity of the detector \cite{Pos_SIPM,Pos_analytical}.

For the purpose of the current work, Compton Cameras can be grouped into two different designs regarding the type of scintillator used: monolithic or segmented crystals. 

For monolithic crystals, the position is continuous within the volume of the active material, with the information being shared between different pixels. In contrast, with detectors that use a segmented material with segments matching the size of the pixel, position information is discrete between the segmentations.

\begin{figure}[h!]
    \centering
    \includegraphics[width=.75\textwidth]{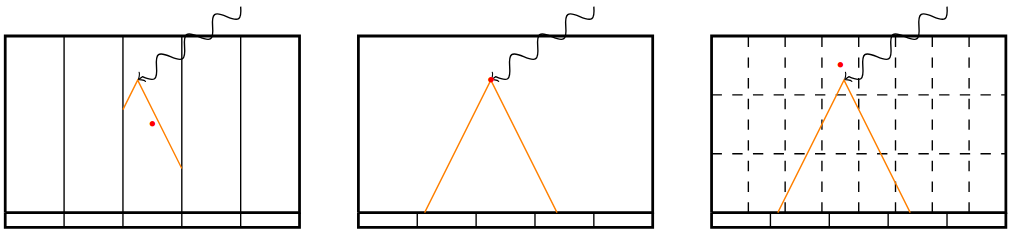}
    \caption{Schematic representation of a segmented (left) and monolithic (center) scintillator crystal used in Compton cameras, and their respective mechanisms of light propagation and collection. On the right, a monolithic scintillator is presented with virtual segmentations imposed into the reconstructed positions of interaction in the center of each segmentation, allowing to treat it as if it was segmented for imaging algorithms. The red dot indicates the reconstructed position of interaction.} \label{TypesCrystal}
\end{figure}

\newpage
\section{Clustering Detections into Events in Monolithic Scintillators}

One of the main challenges of working with Compton Cameras made from large monolithic crystals is related to building events. As the scintillator isn't segmented, the photons arising from the energy deposited by the $\gamma$-ray in the crystal are free to propagate and be captured by any of the pixels in the photomultiplier. As such, one event will result in multiple detections distributed in space and time. This has the added benefit of allowing a position resolution smaller than the pixelation used and to compute the depth of interaction, \cite{Pos_SIPM,Pos_ML}, in exchange for added complexity.

\begin{figure}[h!]
        \centering
        \includegraphics[width=.75\textwidth]{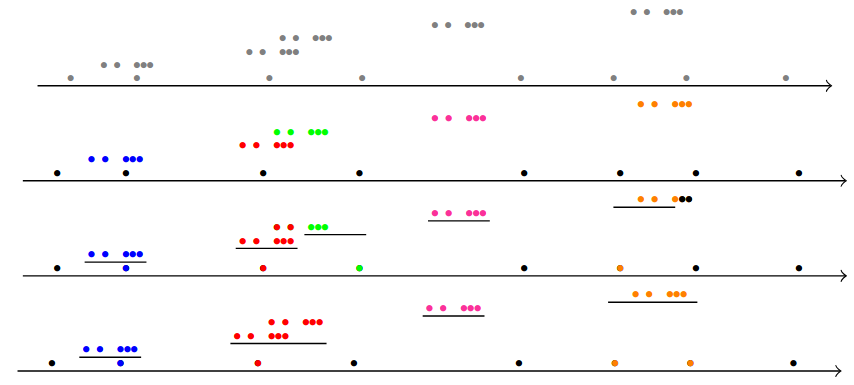}
        \caption{Schematic representation of detections in individual pixels over time due to noise, represented by the bottom points, and true measurements, represented by the other points; of the objective of the clustering algorithm; of the result of using a fixed-length non-overlapping coincidence window; and of using a non-overlapping window of variable length with a maximum step.} \label{AlgoPipelineResult}
    \end{figure}

One common way, \cite{Pos_SIPM}, of clustering the different detections into groups related to the same physical event is by defining the time interval of the detections of an event, through simulation or measurement, and using a fixed-length non-overlapping coincidence window with a threshold in number of detection and total energy deposited. The result of this algorithm is illustrated in \cref{AlgoPipelineResult}.

This algorithm requires that the clusters of detections are identified in order. The signals in the fixed-length moving window will only be selected to correspond to the same event if they meet the thresholds imposed. In this case, the window proceeds to start after the last event of the previous window, being non-overlapping. Otherwise, if the thresholds are not met, the window moves by one signal. This implies that the algorithm cannot be parallelized.

In order to address this limitation, a new algorithm was developed in this work with the goal of enabling parallelization and thus enhance performance.

    The algorithm proposed here is based on a maximum step between consecutive detections, which results in a variable window size, instead of the established fixed-window. By moving the condition for event delimitation from groups of multiple detections to two consecutive detections, the computation becomes vectorizable. The algorithm, which yields a boolean array of the edges indicating the beginning and end of detections that should be clustered into an event, can be described as follows:

    \begin{enumerate}
        \item Calculate the time difference between consecutive detections:\\
        $$\Delta t_\text{i+1,i}=t_\text{i+1}-t_\text{i}$$
        \item Apply a binary mask for time differences less than a maximum time step:\\
        $$M_\text{i+1,i}:=\Delta_\text{i+1,i} < \Delta_\text{max}$$
        \item Apply the XOR operation between consecutive time differences:\\
        $$E_\text{t}:= M_\text{i,i-1} \oplus M_\text{i+1,i}$$
    \end{enumerate}
    
\begin{figure}[h!]
        \centering
        \includegraphics[width=\textwidth]{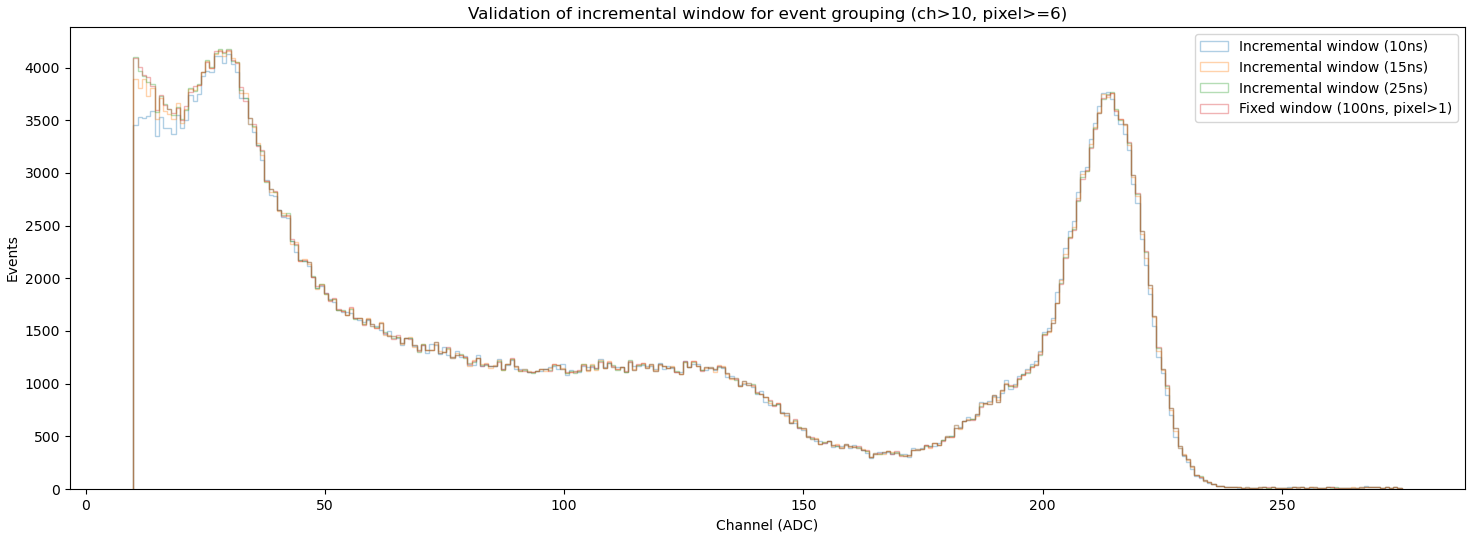}
        \caption{Reconstructed energy deposition spectrum of $^{137}$Cs in channel ADC, measured with a LaCl3 scintillation detector, for the events built from the clustering with a fixed 100ns non-overlapping coincidence window and a non-overlapping window of variable length with a maximum step of 10ns, 15ns and 25ns.} \label{ValidationPipeline}
    \end{figure}

    The proposed algorithm, as made evident by \cref{AlgoPipelineResult}, yields different results than than the common approach, \cite{Pos_SIPM}. However, these fundamental differences bring some added benefits: less susceptible to having a noisy pixel breaking a event, simultaneous events are grouped, which is understood as pileup, instead of breaking events. Furthermore, the energy spectrum of interactions in coincidence of $^{137}Cs$, a standard $\gamma$-ray source widely used for calibration \cite{Pos_SIPM}, present in \cref{ValidationPipeline}, validates the compatibility of the results obtained.

    The algorithm was implemented in SYCL, \cite{SYCL}, a single-source C++ standard for Heterogeneous Computing, using 2 kernels, as the first two steps of the algorithm can be combined.

\newpage
\section{Parallel Back Projection Algorithm for 3D Imaging}

There are several algorithms for Compton imaging as those discussed, for example, in \cite{ImagingAlgos}. However, the present work will focus on the optimization and parallelization of the back projection algorithm for 3D imaging discussed in \cite{BackProjection3D}.

This imaging algorithm calculates, for each $\gamma$-ray, its contribution to each voxel in our space. The contribution to each voxel is calculated based on the probability the $\gamma$-ray came from its center. As such, instead of voxels, it can be interpreted as a set of probes placed in the center of the voxels.

This imaging algorithm is a good candidate for parallelization since the effect of each event in each voxel is independent, in contrast to other approaches such as the SOE, \cite{SOE}.

This algorithm was parallelized using SYCL, \cite{SYCL}. In our implementation, a \texttt{parallel\_for} with $N_\text{Events}\times N_\text{Voxels}$ iterations computes each weighted contribution to a 3D histogram. The contributions are then summed and added to the histogram.

To further improve the performance of the imaging algorithm, several optimizations were applied, which are described below.

\subsubsection{Precomputing values}

Several values that are used in the algorithm can be precomputed and loaded into the memory of the device in order to save computation time. In order to achieve this, discrete positions of interaction in the scatterer, absorber, and of origin in the space need to be known prior to the measurement. As such, if monolithic crystals are being used, virtual segmentations need to be imposed. With these values known it is possible to precompute the geometric Compton angles, based on the three points, as well as the distances between the positions in the scatterer and absorber, and scatterer and voxel center.

\subsubsection{Angular selection of voxels}

The contribution of each interaction to each voxel decreases with the increase of the absolute angular difference between the Compton angle calculated from the three points (geometric) and the one calculated from the energies deposited (energetic). A condition can be placed in order to not calculate the contributions for differences greater than a given angle.

\subsubsection{Spatial optimization}

Lastly, although a cubic shape is the generic choice for the image universe, this is not always necessary. In many cases, if the source being imaged is placed centrally the corners hold little useful information. In many cases a cylindrical, or even spherical shape, suffices. This results in a overall lower number of voxels and, as such, lower number of iterations. For example, a cylindrical shape has about 21.5\% less volume than the cube it is inscribed in, while the sphere has 47.6\% less.

\bigskip

In a preliminary analysis, the results obtained were compatible with the results of the previously established code. A systematic study of the impact of the optimizations previously described in applications in which quantitative dose assessment is central, such as medical and nuclear waste imaging, is warranted.

\section{Conclusion and Future Work}

In this work, the impact of multiple optimizations to the processing chain of Compton Camera detectors is studied. The optimizations focus on the use of parallel computations, offloading to GPU, and avoid unnecessary, as well as, low impact computations. A further study into the performance with systematic benchmarks across hardware architectures and vendors is warranted.

Future work includes exploring curve-fitting and neural networks approaches to reconstruct the position of the energy deposition in the scintillation crystals, continuing the work developed in \cite{Pos_analytical,Pos_ML}, necessary to develop the full processing chain and benchmark the overall effects on performance. Lastly, a study will be conducted into the benefits of the heterogeneous nature of SYCL with the express objective of assessing the feasibility and practicability of fully integrating the functionality of Compton Cameras into a embedded system, with special regard to high counting rate use cases.

\begin{credits}
\subsubsection{\ackname} This study was funded by CSIC (contract PRE2023-IFIC-141, project PID2022-138297NB-C21), and ERC-POC AMA, Grant Agreement 101137646.

\subsubsection{\discintname}
Bernardo Gameiro is, at the time of writing, an Intel Certified Instructor for oneAPI. The other authors have no competing interests to declare that are relevant to the content of this article.

\end{credits}
\bibliographystyle{splncs04}

\end{document}